# Hot carrier diffusion-assisted ideal carrier multiplication in monolayer $MoSe_2$

Joonsoo Kim[1,†], Hong-Guk Min[2,†], Sehwan Park[1,†], Jin Cheol Park[1], Junhyeok Bang[3,*], Youngkuk Kim[2,*], and Ji-Hee Kim[4,*]

[1]Department of Energy Science, Sungkyunkwan University

[2]Department of Physics, Sungkyunkwan University

[3]Department of Physics, Chungbuk National University, Cheongju 28644, Republic of Korea

[4]Department of Physics, Pusan National University, Busan 46241, Republic of Korea

[†]These authors contributed equally to this work.

[*]Corresponding author: kimjihee@pusan.ac.kr (J.-H.K.), youngkuk@skku.edu (Y.K.), jbang@cbnu.ac.kr (J.B.)

**Carrier multiplication (CM), the process of generating multiple charge carriers from a single photon, offers an opportunity to exceed the Shockley-Queisser limit in photovoltaic applications. Despite extensive research, no material has yet achieved ideal CM efficiency, primarily owing to significant energy losses from carrier-lattice scattering. In this study, we demonstrate that monolayer $MoSe_2$ can attain the theoretical maximum CM efficiency permitted by energy-momentum conservation principle, using ultrafast transient absorption spectroscopy. By resolving the scatter-free ballistic transport of hot carriers and validating our findings with first-principles calculations, we identify the cornerstone of optimal CM in monolayer $MoSe_2$: superior hot-carrier dynamics characterized by suppressed energy dissipation via minimized carrier-lattice scattering, and the availability of abundant CM pathways facilitated by $2E_g$ band nesting. Comparative analysis with bulk $MoSe_2$ further emphasizes the enhanced CM efficiency in the monolayer, attributed by superior hot-carrier diffusion and access to additional CM pathways. These results position monolayer $MoSe_2$ as a promising candidate for high-performance optoelectronic applications, providing a robust platform for next-generation energy conversion technologies.**

## 1. Introduction

Carrier multiplication (CM) occurs when a single high-energy photon generates multiple free electron-hole pairs in a material, surpassing twice the bandgap energy and thus mitigating energy loss as heat. This mechanism is fundamentally distinct from multiple exciton generation (MEG) process, which typically involves the formation of bound multiexcitons rather than free carrier[1, 2]. Utilizing CM, photovoltaic systems have the potential to improve energy conversion efficiency significantly. The benchmark for such enhancement is the "ideal CM efficiency", manifested as a step-like increase in quantum yield (QY) at the threshold of twice the band gap ($2E_g$). This is distinct from the traditional Shockley-Queisser limit[3, 4]. Despite intensive research across a range of materials, including bulk and quasi-two-dimensional thin films of silicon, germanium, graphene, carbon nanotube, lead-chalcogenides, cadmium-chalcogenides, and perovskites[5-14], the highest CM efficiencies achieved to date fall short of this quantum limit, underscoring the challenge of reaching the ideal.

Achieving the quantum limit of the ideal CM efficiency requires meticulous adjustment of fundamental material properties (**Figure 1a**). Initially, to lower the CM threshold energy ($E_{th}$) to $2E_g$, it is essential to have active CM channels at $2E_g$ excitation. However, such channels are typically obstructed by stringent energy-momentum conservation constraints[15]. Secondly, achieving a step-like increase in QY consistent with ideal CM, illustrated by the line of $\eta_{CM}$ = 100% in **Figure 1a**, necessitates that the CM rate exceeds the rate of carrier-lattice scattering losses during thermalization[4]. Notably, recent studies have demonstrated that new emerging van der Waals layered thin films can achieve high CM efficiency ($\eta_{CM}$ = 95%) with the threshold energy of 2.05 $E_g$ via optical and electrical techniques[16-20]. Despite these advancements, the integrated number of carriers produced by this CM process remain around 58%p fewer than those in the ideal CM scenario (see **Figure S1** in **Supporting information**).

In this work, we demonstrate that an atomically thin $MoSe_2$ monolayer enables the ideal CM efficiency. By utilizing ultrafast transient absorption spectroscopy, we observe the onset of CM at $2E_g$ with a quantum yield of 2.0. Our first-principles calculations corroborate the observed ideal CM in the $MoSe_2$ monolayer, revealing an electronic band structure conducive to abundant CM channels at $2E_g$, a phenomenon we attribute to $2E_g$ band nesting. Furthermore, femtosecond spatiotemporal spectroscopy measurements substantiate the feasibility of achieving ideal CM, unveiling exceptional hot-carrier transport properties within the monolayer, characterized by significantly enhanced diffusivity unique to this system. Our demonstration of ideal CM, underpinned by its fundamental principles, positions the $MoSe_2$ monolayer as a promising candidate for future electronic devices and energy harvesting applications.

## 2. Results and Discussion

We synthesized an atomically thin monolayer of 2H-$MoSe_2$ using chemical vapor deposition (see **Figure S2** in **Supporting information** for sample characteristic). The optical properties of the $MoSe_2$ monolayer were confirmed through steady-state absorption and photoluminescence (PL) spectroscopy. The steady-state absorption spectrum revealed the well-documented $A$ and $B$ excitons of $MoSe_2$[21], manifesting as peaks at 1.58 eV and 1.79 eV, respectively (see **Figure S3** in **Supporting information**). Additionally, the $C$ exciton appeared as a broad peak at higher energies, indicative of band nesting resulting from a shift in the valence band[22]. We point out that this type of band nesting, related to the $C$ exciton, differs from the $2E_g$ band nesting we discuss later, which involves a shift in the valence band by $2E_g$ and plays a critical role in facilitating CM channels. The distinct peak at 1.54 eV (805 nm) in the PL spectrum is consistent with the optical bandgap of monolayer $MoSe_2$ [23-25].

Employing ultrafast transient absorption (TA) spectroscopy, we have uncovered compelling evidence of ideal CM in monolayer $MoSe_2$, marked by distinct spectral signatures and a quantitative relationship between absorption changes and photon density. A key observation emerged when monitoring the kinetics of differential absorption $-\Delta\mathrm{A}(E,t)$ at probe energies of $E_{\mathrm{probe}}$ = 1.58 eV (A exciton), as illustrated in **Figure 1b**: a sudden two-fold increase in the maximum differential absorption ($|\Delta\mathrm{A}(E)|_{\max}$) occurred when the pump energy exceeded $2E_g$ with the absorbed photon density at 1.5 $\times 10^{12}$ photons $cm^{-2}$. As the intensity of the TA signal is linearly proportional to the carrier density, this observation indicates the generation of twice the number of carriers. The pseudo-color TA map (Figure S4) and decay dynamics (Note 1 and Figure S5 in the Supporting Information) further confirm our observation. This significant finding, together with the detection of two additional CM indicators—a transient Stark shift and delayed build-up time (see **Note 2** and **Figure S6-11** in **Supporting information** for detailed analysis)—highlights the exceptional CM efficiency in this system.

These observations directly demonstrate the ideal CM efficiency achieved in monolayer $MoSe_2$. To minimize nonlinear effects, including many-body interactions, the $MoSe_2$ was excited with a sufficiently low photon density (see **Figure S9** and **S12** in **Supporting information**). As shown in **Figure 1c,** $|\Delta\mathrm{A}(E)|_{\max}$ increases linearly with absorbed photon density across all pump energies. For pump energies below $2E_g$, the QY slope is one, indicating that a single carrier is generated per photon (see **Note 3** in **Supporting information** for details). When the pump energy exceeds $2E_g$, the QY doubles to two, signifying the excitation of two carriers per photon. Furthermore, **Figure 1d** shows that QY as a function of photon energy exhibits a sharp, quantized jump at the threshold energy of $2E_g$, confirming the ideal CM efficiency in monolayer $MoSe_2$. This behavior is in stark contrast to the gradual increases in QY observed in bulk $MoSe_2$, which deviate significantly from the ideal CM efficiency. Notably, the QY curve for monolayer $MoSe_2$ closely follows the simulated quantum-limit curve, as indicated by the black solid line in **Figure 1d** (see **Note 4** in the **Supporting information**). The reproducibility of ideal CM in monolayer $MoSe_2$ is confirmed by consistent results across multiple measurements (see **Figure S13** in Supporting information). This alignment demonstrates that monolayer $MoSe_2$ achieves ideal 100% CM efficiency ($\eta_{CM}$), whereas the bulk counterpart and all previously studied materials[13, 14, 17] fall short.

Having demonstrated the ideal CM efficiency in monolayer $MoSe_2$, we now turn to our first-principles calculations to uncover the electronic structure responsible for this exceptionally high efficiency. Using a previously developed method [26], we count the number of possible CM channels $N(I_1)$ for a given initial state of a carrier $I_1$ generated by a photon with $E = 2E_g$ (threshold energy) by examining the energy band structure across the entire Brillouin zone (see **Note 5** and **Figure S14-15** in **Supporting information** for the detailed methods and analysis). As shown in **Figure 2a**, the calculated $N(I_1)$ yields non-zero values in the conduction bands around the $K$ and $K'$ points. These calculations indicate that numerous CM channels exist through which hot electrons excited by photons with $E = 2E_g$ in the vicinity of the $K$ and $K'$ points can contribute to the CM process, thereby theoretically supporting the threshold energy of $2E_g$ observed in monolayer $MoSe_2$. We also note that our calculation is consistent with the results for bulk $MoSe_2$. It is found that the CM channels for carriers excited at $2E_g$ are entirely blocked in the bulk system due to energy-momentum conservation laws (see **Figures S16 and S17** in Supporting information and **Note 5** for the calculations of CM channels in bulk $MoSe_2$), which aligns with the observed threshold energy of $E = 2.05\ E_g$ for bulk $MoSe_2$, as shown in **Figure 1d**.

The presence of $2E_g$ CM channels in monolayer $MoSe_2$ is linked to the generic electronic structure of monolayer transition metal dichalcogenides (TMDs), which is characterized by $2E_g$ band nesting and valley symmetry. Specifically, near the $K$ (or $K'$) point, the highest occupied bands and the third lowest unoccupied bands align closely when shifted by $2E_g$ (**Figure 2c**), referred to as $2E_g$ band nesting. The alignment is a common feature in monolayer TMDs, arising from their crystal symmetry, which results in an approximately equal splitting of the hybridized $d_{x^2-y^2}$ and $d_{xy}$ orbitals. The $2E_g$ nesting is absent in the valence bands, thus the $2E_g$ CM channels from the hole side are blocked in monolayer $MoSe_2$ (see **Note 6** and **Figures S18** in **Supporting information**). As detailed in **Figure S16** in **Supporting information**, these nested bands facilitate various intra-valley (see **Figure 2b**) and inter-valley CM channels near the $K$ and $K'$ points. Additionally, the valley degeneracy of the monolayer ensures that the number of intra-valley CM channels equals the number of inter-valley

CM channels, supported by local inversion and time-reversal symmetry, collectively referred to here as valley symmetries. These symmetries enhance the number of CM channels for $2E_g$ excitation, where a single intra-valley CM channel at $K$ or $K'$ leads to the creation of two additional inter-valley CM channels between $K$ and $K'$ (see **Note 5** in **Supporting information** for detailed analysis). Thus, we believe that the enhancement of CM efficiency observed in monolayer $MoSe_2$ may be expected in other monolayer TMDs as well.

Having established the existence of $2E_g$ CM channels in monolayer $MoSe_2$, we now focus on the transport properties of hot carriers, which are essential for realizing ideal CM efficiency. Investigating hot carrier transport on sub-picosecond timescales is important, as the high mobility of these hot carriers can delay Auger recombination and thermalization by facilitating the rapid spatial separation of carriers, thereby enhancing CM efficiency[27, 28]. After the CM process, the increased population of carriers raises the likelihood of efficiency-reducing events, such as Auger recombination and thermalization. These competing processes occur alongside CM and can significantly reduce its efficiency. Furthermore, TMDs like $MoSe_2$ have relatively high electron-hole binding energies[23, 29], which further increase the rate of Auger recombination[30]. In the following sections, we will demonstrate that the high mobility of hot carriers in monolayer $MoSe_2$ effectively delays these losses, thus supporting sustained CM efficiency.

Employing femtosecond spatiotemporal transmission absorption microscopy, we unveil the capacity of monolayer $MoSe_2$ for enhanced spatial separation and delayed thermalization of hot carriers (for detailed experimental methods, see **Note 7** and **Figure S19a** in **Supporting information**). **Figure 3a** presents the spatiotemporal dynamics of excited carriers in monolayer, shedding light on hot-carrier diffusion processes (detailed analysis provided in **Note 8**). Unlike their bulk counterparts, hot carriers in the monolayer demonstrate immediate spatial separation and exhibit significantly extended lifetimes, a comparison starkly evident in the juxtaposed top and bottom panels of **Figure 3a**. This differential behavior is further highlighted (see **Figure S19c** in **Supporting information)**, in which diffusion profiles are traced within a $\pm 2\ \mu$m range. Remarkably, within a sub-picosecond duration, hot carriers in the monolayer are shown to spread across this range, showcasing rapid and efficient spatiotemporal separation.

Our examination of the excited carrier dynamics in the monolayer $MoSe_2$ reveals a pronounced increase within the first two picoseconds. We obtain squared width broadening $\Delta\sigma^2(t) = \sigma^2(t) - \sigma^2(0)$, where $\sigma(t)$ is the Gaussian width as a function of time, from the measured profiles (see **Figure S19c** in **Supporting information)**. This dynamic behavior is fitted by a power function $\Delta\sigma^2(t) = Dt^{\alpha}$ with $\alpha = 1$, from which we calculate a diffusion coefficient ($D_{\text{mono}} = \Delta\sigma^2/2t$), showing a maximum diffusion rate of 1.0 × $10^4$ $cm^2$ $s^{-1}$ (**Figure 3b,** see **Figure S20** in **Supporting information** for details on long-range diffusion). Remarkably, this rate surpasses that for other CM materials[31, 32] and even Au[33] by one to two orders of magnitude, evidencing the exceptional hot-carrier expansion capability of the monolayer. The minimal influence of carrier–carrier scattering on this diffusivity indicates that the enhanced hot-carrier expansion characteristic of the monolayer is likely responsible for its reduced carrier–lattice scattering rates. Highlighting the distinct advantage of monolayer $MoSe_2$ in hot-carrier transport dynamics, these findings point to its superior potential for efficient CM, markedly setting it apart from other materials. After the rapid hot-carrier diffusion phase, a subsequent negative diffusion phase is observed (**Figure 3b**) in monolayer, attributed to the formation of excitons from hot-carrier states. This observation is consistent with prior results (see **Note 8** in the **Supporting Information** for detailed analysis)[31, 34]

In contrast to the monolayer, our measurements reveal a significantly higher carrier-lattice scattering rate in bulk $MoSe_2$, as evidenced by the slower hot-carrier expansion depicted in **Figure 3b**. For the bulk material, $\Delta\sigma^2(t)$ showed a steady increase only up to $t$ = 5 ps, resulting in a hot-carrier diffusion coefficient of $D_{\text{bulk}}$ = 2.6 × $10^3$ $cm^2$ $s^{-1}$ in the initial phase—substantially lower by an order of magnitude compared to the monolayer. This discrepancy is attributed to the bulk carriers to occupy in-plane dimensions with less confinement out-of-plane, offering a broader array of states for thermal excitation than in the monolayer. In contrast, in 2D systems like the monolayer $MoSe_2$, rapid hot carrier expansion within the sub-picosecond range effectively reduces the density of excited carriers in the spatial domain[27, 35]. Such dynamics are crucial for enhancing CM efficiency in monolayer structures by preventing spatial congestion of the excited electrons, thereby facilitating a more efficient CM process.

To shed light on the diffusion mechanism of hot carriers, we analyzed their initial spreading by fitting it to the power function $\Delta\sigma^2(t) = Dt^{\beta}$, where the $\beta$ represents transport exponent determining the nature of carrier spreading. A $\beta$ value of 1 denotes classical diffusion, typified by the scattering motion of particles, whereas $\beta = 2$ suggests ballistic motion, characterized by scattering-free carrier transfer[32, 36, 37]. In **Figure 3c**, we display the power fit as a function of delay time for the monolayer, revealing that carriers propagate with $\beta = 2$ up to 0.7 ps (indicated by a red dot), implying that hot carriers experience diffusion predominantly through ballistic transport initially[32, 38] (see **Note 9** and **Figure S21** in **Supporting information** for detailed fitting method). However, beyond this sub-picosecond regime, the fit begins to deviate from $\beta = 2$ (marked by a green dot), transitioning towards linear diffusion ($\beta = 1$), as denoted by a blue dot in **Figure 3c**.

To further elucidate ballistic transport, we drew a correlation between the excess energy of photons and energy conservation[37]. During photoexcitation, this excess energy is directly converted into kinetic energy, as described by the equation of $E_{ex} = \frac{1}{2} m^* v_B^2$, where $v_B$ is ballistic velocity, $E_{ex}$ is excess energy of photon, and $m^*$ is effective mass. For the monolayer, the ballistic velocity by excess energy is computed to be $6.6 \times 10^5\ m/s$, which is comparable to the experimentally obtained maximum hot-carrier velocity ($\partial\sigma/\ \partial t|_{max}$) of $7.2 \times 10^5\ m/s$, supporting the presence of ballistic transport in this system.

Building on the observed ballistic motion and subsequent hot-carrier expansion in $MoSe_2$, we delve into the physical processes underpinning carrier multiplication (CM) and their spatial distribution over time, as depicted in **Figure 3d**. Following photoexcitation, carriers are instantaneously excited to match the beam size of the system. Within the first 0.7 ps, these carriers undergo scatter-free ballistic movement due to their excess energy, enabling rapid diffusion at rates surpassing $10^4$ $cm^2/s$. This swift expansion aids in spatially separating the accumulated carriers, significantly mitigating intra-band scattering as well as carrier-carrier annihilation.

It is critical to acknowledge the competitive dynamics between CM and Auger recombination[6]. In the absence of ballistic transport, as depicted in **Figure 3d**, the CM process can lead to the accumulation of carriers in both energy and spatial domains. This accumulation increases the likelihood of carrier-carrier interactions, which in turn elevates the rate of carrier-carrier annihilation and Auger recombination. As a result, the observable carrier density diminishes, effectively negating the additional carrier generation achieved through CM. In contrast, when ballistic transport occurs concurrently with CM, the system potentially benefits from the spatial separation of carriers, which significantly reduces the rate of Auger recombination and carrier-carrier annihilation[27, 35]. Given the elevated kinetic energy in the monolayer, we anticipate that the CM in monolayer $MoSe_2$ surpasses that of its bulk counterpart.

The realization of ideal CM in monolayer $MoSe_2$ represents a significant milestone in the field of optoelectronics. To contextualize our finding within the broader landscape of CM research, we examined the CM efficiency across various material dimensions, as depicted in **Figure 4**. Each point on the graph represents the CM efficiency of a specific material, with detailed data provided in **Table S4** of the **Supporting Information**. The solid line indicates the average CM efficiency for each dimensional category. Quantum dots (0D) are well-established for their high CM efficiencies, with recent advances in perovskite quantum dots achieving over 87% efficiency through doping to reduce hot-carrier trapping, resulting in an internal quantum efficiency of 120%[14]. Nevertheless, 2D materials, including monolayer $MoSe_2$, exhibit even higher CM conversion efficiencies compared to same substance in other dimensional forms. Moreover, the 2D category also shows the highest average efficiency overall. his exceptional performance, as underscored by Aerts et al., is primarily driven by the pronounced quantum confinement in the out-of-plane direction, coupled with a high density of states in the in-plane direction—factors that are critical for enhancing CM efficiency in 2D materials[11]. Moreover, 2D nanosheets, such as PbS, PbSe, and perovskites, typically form through polycrystalline structures or molecular interconnections which introduce discontinuities in the density of states. These discontinuities promote scattering centers, increasing carrier trapping and ultimately reducing CM efficiency. In contrast, van der Waals materials like TMD and black phosphorus (BP) consist of covalently bonded atomic layers in the in-plane direction, maintaining continuous in-plane transport while preserving strong quantum confinement effects. Specifically, our investigation reveals that monolayer $MoSe_2$ capitalizes on a unique combination of band nesting effects and ballistic transport properties. These factors collectively suppress energy loss mechanisms which are prevalent in other material systems. Consequently, monolayer $MoSe_2$ achieves unparalleled CM efficiency, outperforming bulk and lower-dimensional counterparts.

The exceptional CM efficiency observed in monolayer $MoSe_2$ holds great potential for advanced optoelectronic applications. The comparative analysis (**Figure 4**) reveals that 2D systems, and especially monolayer $MoSe_2$ in particular, outperform 0D and 1D counterparts. The efficient CM conversion is particularly beneficial for photovoltaic devices, as it minimizes thermalization losses and can help overcome the Shockley–Queisser limit[17, 20]. The intrinsic $2E_g$ band nesting and valley symmetry (**Figure 2c**) provide multiple CM pathways that could enhance electroluminescence efficiency in light-emitting devices. In addition, the observed ballistic transport ($10^4$ $cm^2/s$, **Figure 3b**) and rapid hot-carrier diffusion enable efficient carrier extraction while minimizing recombination losses (Figure 3d), making monolayer an excellent candidate for ultrafast photodetectors. This observation establishes monolayer $MoSe_2$ as a promising platform for next-generation energy-harvesting, quantum optoelectronic, and photonic technologies (see Note 10 in Supporting Information for further details on scalable synthesis, device integration, and long-term stability strategies).

While our results demonstrate exceptional QY behavior in monolayer $MoSe_2$ (**Figure 1d**), the microscopic mechanism underlying this CM remains a subject of debate, with potential contributions from either CM or multiple exciton generation (MEG) in $MoSe_2$. Based on our findings, we suggest that the observed efficiency is more likely attributable to CM rather than MEG. MEG typically relies on exciton-exciton interactions within bound excitonic states to generate additional carriers[9, 13]. In contrast, the dynamics observed within the CM timescale strongly suggest a process dominated by free carrier generation and transport[31, 39-41]. The rapid hot carrier diffusion behavior presented in **Figure 3c**, including ballistic transport, strongly supports this hypothesis. As discussed in **Note 8** in **Supporting information**, the negative diffusion in monolayer $MoSe_2$ (**Figure 3b**) is attributed to hot carrier cooling to form exciton formation, a process that occurs on a much longer time scale than CM[31, 39-41]. Furthermore, the rapid diffusion of hot carriers reduces the likelihood of exciton-exciton interactions, supporting the interpretation of a free carrier-driven CM process in monolayer $MoSe_2$. We noted that although our study observed a two-fold increase in the number of carriers induced by CM, the transition from a single carrier to multiple carriers was not directly resolved within the temporal resolution of our system. This suggests that the process occurs on a timescale shorter than a few tens of femtoseconds. To fully unravel the underlying mechanism of the ideal CM, techniques with higher temporal resolution, such as femtosecond time- and angle-resolved photoemission spectroscopy (tr-ARPES), could provide critical insights into the formation and subsequent multiplication, thereby offering the profound understanding of the intricate processes[7, 42].

Lastly, the ideal CM that we have achieved signifies a substantial increase in hot carrier generation. By postulating a uniform initial energy distribution of carriers, we can assess CM efficiency using the ratio $R_{CM}$, defined as $R_{CM} = (N_{CM} - N_0)/N_0 \times 100\%$, where $N_{CM}$ and $N_0$ represent the number of carriers generated within the energy range from $2E_g$ to $3E_g$ with and without CM, respectively (see **Figure S1** in **supporting information**). This ratio quantifies the increase in carrier generation attributable to CM compared to the scenario without CM. As illustrated in **Figure 1a**, van der Waals layered thin films currently reach an $R_{CM}$ of only 42%, falling short of the half of the enhancement goal[16]. This shortfall underscores significant opportunities for further optimization of $R_{CM}$ toward the ideal CM ($R_{CM} = 100\%$). The strides made in realizing ideal CM in this study markedly influence hot carrier generation, opening new avenues for optimizing and harnessing CM efficiency.

## Conclusions

In this work, we have demonstrated the realization of ideal CM efficiency in monolayer $MoSe_2$, achieving a quantum yield of 2.0 at the threshold energy of $2E_g$. Our findings, supported by ultrafast transient absorption spectroscopy and first-principles calculations, underscore the unique electronic properties of monolayer $MoSe_2$ that enable efficient CM, specifically its $2E_g$ band nesting and valley symmetry, which facilitate the formation of abundant CM channels. Furthermore, our spatiotemporal spectroscopy revealed superior hot-carrier transport in monolayer $MoSe_2$, including rapid ballistic transport and reduced carrier-lattice scattering, which further contribute to the observed CM efficiency by minimizing recombination losses. Our results highlight the potential of two-dimensional materials in overcoming the limitations of bulk semiconductors and pave the way for further research aimed at optimizing CM processes in other materials. The demonstration of ideal CM efficiency, coupled with long-range ballistic transport, marks a pivotal advancement in the optimization of energy conversion processes. This work not only advances the fundamental understanding of CM in 2D materials but also provides a promising avenue for the development of next-generation optoelectronic devices, with the potential to surpass conventional photovoltaic efficiency limits.

***Experimental section***

**Sample synthesis and characterization**

The $MoSe_2$ film sample was grown by an atmospheric chemical vapor deposition system. The liquid precursor for $MoSe_2$ film was prepared by mixing following aqueous solutions in a ratio of A:B:C=1:3:0.1. (A) 0.1g of sodium molybdate ($Na_2MoO_4$, Sigma-Aldrich, 737860) dissolved in 10 ml of deionized (D.I.) water, (B) D.I. water (c) OptiPrep density gradient medium (Iodixanol, Sigma-Aldrich, D1556).

The prepared liquid precursor was spin-coated onto a 300 nm $SiO_2$/Si substrate at 3000 rpm for 60 s. Each 0.4 g of selenium pellets (Sigma-Aldrich, 204307) and the precursor-coated substrate were separately introduced in a 2-inch two-zone tube furnace. The selenium zone was heated to 380 oC at 50 oC min-1, while the substrate zone was heated to 750 oC at 100 oC min-1 with 600 sccm flow of nitrogen gas (99.9999% purity) and 4 sccm flow of hydrogen gas (99.999% purity). After 10 min of growth, both zones were opened and naturally cooled down to room temperature.

The as-grown $MoSe_2$ film on $SiO_2$/Si was transfer onto 2 mm-thickness quartz substrate, respectively. The steady-states absorption spectrum is obtained using spectrometer (Stellarnet, USA). The tapping mode AFM images are presented in **Figure S2** in **Supporting information** and were measured using an XE-7 microscope (Park Systems, Korea). Raman spectroscopy was carried out using an NTEGRA SPECTRA (NT-MDT, Netherlands).

**Transient absorption spectroscopy**

A pulse laser was generated by a mode-locked Ti:Sapphire amplifier (COHERENT, USA) with sub-25-fs pulse duration and 1 kHz repetition rate. The central wavelength of the pulse was 800 nm, divided into ~3:1 by a beam splitter. The intense pulse was applied to an optical parameter amplifier (Light Conversion, Lithuania) with tuning from 250 to 15,000 nm, and was used as a pump beam. We used pump energies of 3.54, 3.35, 3.26, 3.18, 3.10, 3.06, 2.76, 2.48, 2.07 and 1.97 eV to generate photoexcited carriers below and above the $2E_g$ range. The absorbed photon density was controlled to within a range of $5 \times 10^{11}$ to $4 \times 10^{12}$ photons $cm^{-2}$. The weak pulse was focused on a nonlinear crystal to generate a white-light continuum (450~800 nm) that was used as a probe beam. The white light was divided by a beam splitter; one beam passed through the sample to detect the transient absorption signal (probe), and the other was used to reduce noise levels (reference). The signal and reference signals were collected by a spectrometer (Helios Fire, Ultrafast Systems, USA) with 2-nm spectral resolution. The delay stage controlled the probe beam path for time delay with the pump pulse.

The transient absorption signal was obtained by measuring the differential absorption, $\Delta\mathrm{A}(E, t)$, with and without pump excitation for a given probe energy $E$ and time delay $t$, described as:

$$\Delta\mathrm{A}(E, t) = \mathrm{A}_{pump-on}(E, t) - \mathrm{A}_{pump-off}(E, 0).$$

The carrier distribution in the spectral range could then be analyzed using the broadband probe and spectrometer. The relative time delay between the pump and probe was used to track carrier population dynamics, such as the recombination process. To this end, we obtained a three-dimensional spectrum and time-resolved dataset. In the spectral analysis, after the pump excited the carriers ($t > 0$), the carriers rapidly thermalized and occupied the conduction band edge. The filled band induced reduction of absorption ($\Delta\mathrm{A} < 0$) at bandgap energy due to Pauli blocking or band-filling effects, indicated as photobleaching. Excited carriers promoted increased absorption to an upper state ($\Delta\mathrm{A} > 0$), called photoinduced absorption.

**Femtosecond spatiotemporal transient absorption microscopy (TAM) technique**

The TAM experiment employed the same laser source (Ti:Sapphire amplifier) used in our carrier multiplication experiments. The pump wavelength was tuned to 600 nm using an optical parameter amplifier (Light Conversion, Lithuania) with a photon density of approximately 1 x $10^{14}$ photons $\mathrm{cm}^{-2}$. The single wavelength pump and the broadband probe pulse in the visible range passed through an objective lens (Nikon, TU Plan ELWD, ×50, 0.60 NA) and were focused on the sample with a 2-$\mu$m spot size at 600 nm. The probe light passing through the sample was collimated by another objective lens (Nikon, TU Plan ELWD, ×50, 0.60 NA) and detected by the spectrometer (Helios, Ultrafast system, USA).

**Density functional theory (DFT) calculation**

First-principles calculations based on DFT were performed using the *Quantum Espresso* package[43]. We considered spin-orbit coupling using fully relativistic pseudopotentials, generated by the *OPIUM* code[44, 45]. The Perdew-Burke-Ernzerhof generalized gradient approximation (PBE-GGA)[46] was employed to approximate the exchange-correlation energies. The plane-waves basis was generated under the kinetic energy cutoff of 60 Ry. For atomic and electronic structure calculations, we used the Monkhorst-Pack sampling[47] of $\boldsymbol{k}$ points from the 12×12×1-grid and 12×12×4-grid of the monolayer and bulk $MoSe_2$ first Brillouin Zone (BZ), respectively. The atomic structure of monolayer $MoSe_2$ was obtained by minimizing the total energy with a fixed interlayer distance to 16.0 Å, sufficiently large enough to mimic the vacuum for the monolayer. These calculations led to the in-plane lattice constant $a$ = 3.30 Å. The internal coordinates were fully relaxed for given lattice parameters, which led to the vertical distance between the nearest Mo and Se layers $z \sim 1.67$ Å, which agrees well with the literature[48]. For the bulk $MoSe_2$ structure, the atomic structure was fully relaxed until the Hellmann-Feynman forces were fully reduced below $10^{-6}$ Ry/Bohr, leading to the bulk lattice constants $a \sim 3.30$ Å and $c \sim 12.91$ Å in good agreement with previous studies[49, 50].

The Wannier Hamiltonian was generated using the Wannier90 package[51]. We used Mo $d$ orbitals and Se $p$ orbitals as initial projectors to describe the energy bands near the Fermi level $E_F$. The Wannier Hamiltonian reproduced the DFT bands in the frozen window of $-4.99$ eV $< E - E_F <$ 3.51 eV and $-3.95$ eV $< E - E_F <$ 3.55 eV for the monolayer and bulk $MoSe_2$, respectively. We employed the constant shift approach to model the experimental direct and indirect band gap of the monolayer and bulk $MoSe_2$, respectively. The monolayer conduction bands were shifted by 0.14 eV, while the bulk conduction bands were shifted by 0.33 eV. These shifts result in experimental band gaps of 1.19 eV and 1.54 eV for the bulk and monolayer, respectively.

While advanced methods such as GW with Bethe-Salpeter formalism are necessary to reproduce the exact band gap, we adopted this constant shift approach to effectively align the calculated band gaps with experimental results. This method is particularly suitable for demonstrating the abundance of $2E_g$ channels originating from band nesting in monolayer $MoSe_2$, which is the focus of this study.

While more advanced schemes, such as GW and HSE calculations, may improve the accuracy of band gaps and the overall shape of conduction bands, the $2E_g$ band nesting responsible for CM remains largely unchanged across different computational approaches. Since CM relies on the presence of these nesting features rather than the fine details of the conduction band dispersion, the PBE+constant shift scheme remains an appropriate and effective choice for demonstrating $2E_g$ CM channels.

### *Competing interests*

The authors declare no competing interests.

### *Data availability*

The data supporting this article have been included as part of the Supplementary Information.

### *Acknowledgments*

This work was supported by the National Research Foundation of Korea (NRF) grant funded by the Korea government(MSIT) (RS-2024-00356964 and RS-2024-00406152). Y.K. acknowledges support from the NRF Grants (NRF-2021R1A2C1013871). Computational support is provided by the KISTI (KSC-2021-CRE-0116). J.B. acknowledges support from the NRF Grants (NRF-2022R1F1A1073305 and NRF-2023R1A2C1006433).

### *Author contributions*

J.-H.K. conceived the idea and designed the project. J.K. performed time-resolved measurements. H.-G.M, J.B., and Y.K. contributed to the Density functional theory calculations. S.P. and J.C.P. synthesized monolayer and bulk transition metal dichalcogenides samples. J.K. and S.P. characterized the samples. J.-H.K., J.K., J.B and Y.K. interpreted the results, and wrote the manuscript. All authors contributed to the manuscript.

## Additional information

### *Associated contents*

Comprehensive evidence of carrier multiplication (CM), including detailed evaluation of quantum yield and quantification of CM efficiency. Additional sections cover CM calculation methodologies using DFT, spatiotemporal carrier dynamics, carrier diffusion, and ballistic transport in $MoSe_2$.

The online version contains supplementary material available at

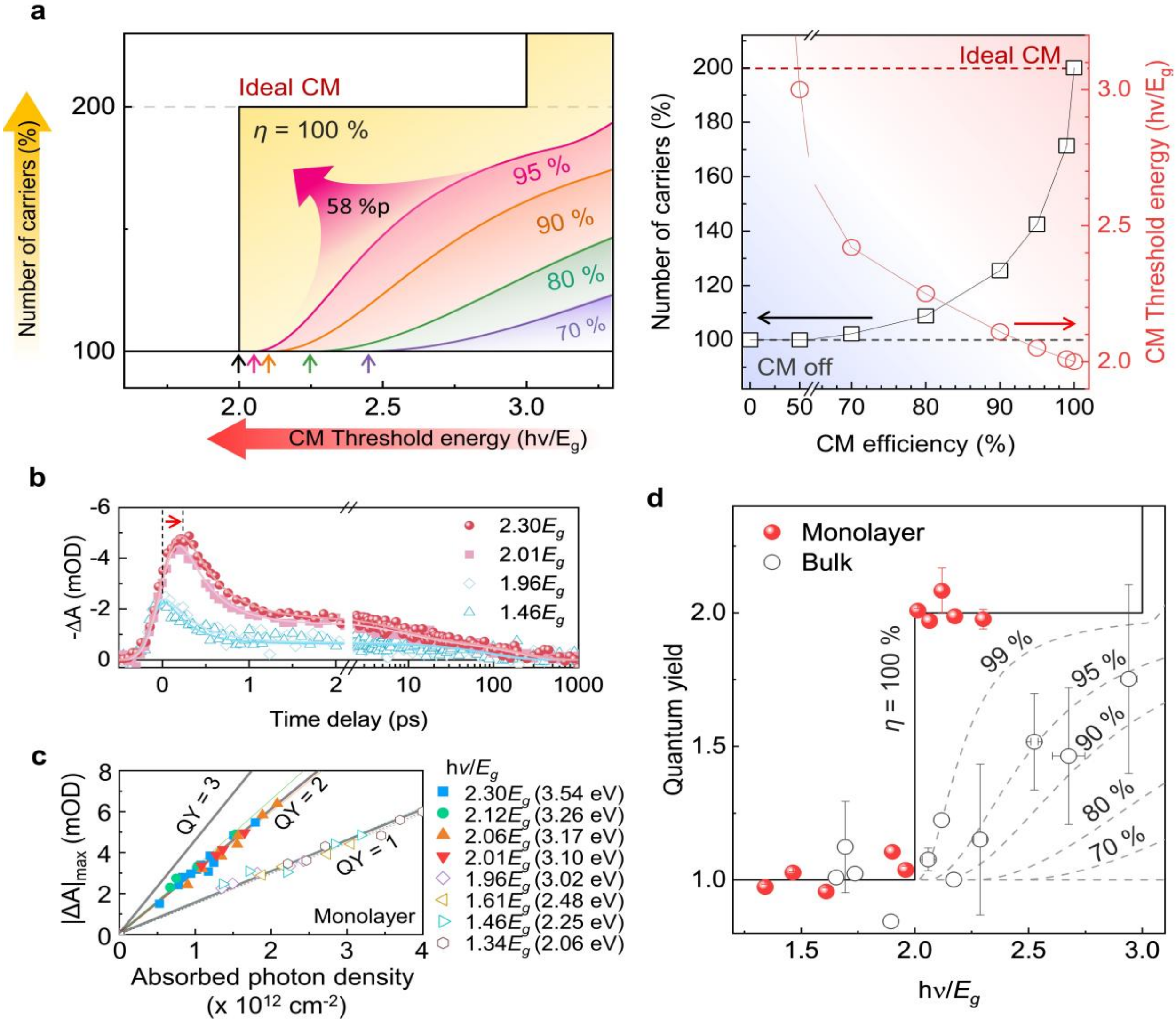

**Figure 1**. **Carrier multiplication (CM) in monolayer $MoSe_2$.** a) The number of generated carriers per photon is plotted, illustrating different CM conversion efficiencies (70%–95%). The CM threshold energy ($E_{th}$) decreases with increasing CM efficiency, and the integrated number of carriers rises by 58%p as CM efficiency increases from 95 to 100%. The right inset depicted the correlation between the integrated carrier generation in the range of $2E_g < h\nu < 3E_g$ and the threshold energy for different CM efficiencies. b) Photobleaching dynamics of $|\Delta A(E,t)|$ at $A$ exciton energy are plotted for various excitation energies with the same photon density (1.5 x $10^{12}$ $cm^{-2}$). c) The maximum intensity of $\Delta A$ is derived from carrier dynamics as a function of absorbed photon density for different excitation energies. The gray lines represent the quantum yield (QY), quantifying the rate of carrier generation. d) QY is plotted against pump energy, normalized by bandgap ($E_g$), for monolayer $MoSe_2$ (red circles) and bulk $MoSe_2$ (open circles). Error bars reflect the uncertainty in excitation energy and experiment repeatability. The black solid and dashed lines depict simulation results corresponding to each CM efficiency.

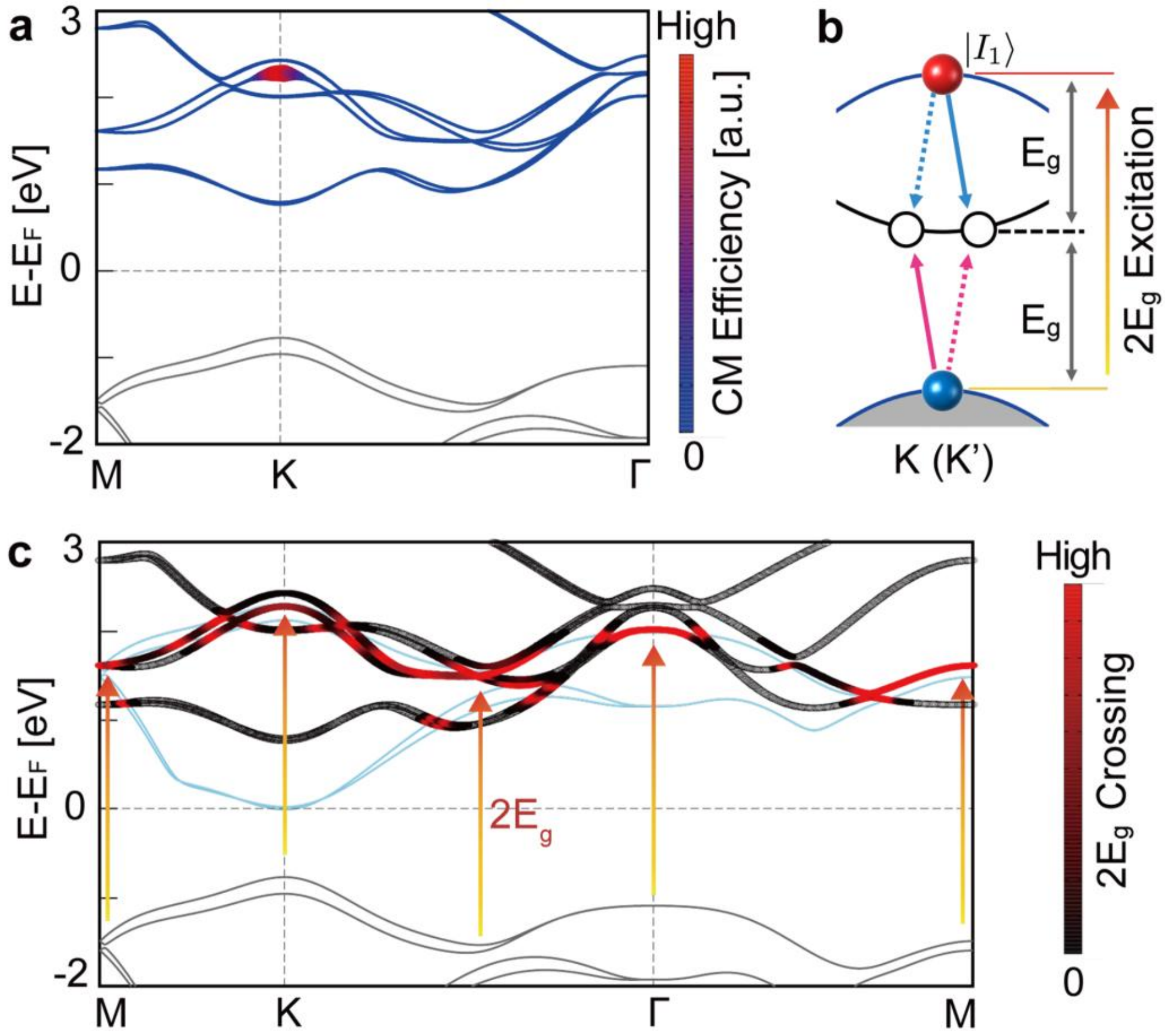

**Figure 2**. **DFT calculations of CM efficiency in monolayer and bulk MoSe2.**

a) DFT calculations of the CM channels in monolayer $MoSe_2$. The CM channel density $N(I_1)$ is calculated for given conduction states $I_1$ in monolayer $MoSe_2$. Regions depicted in red, characterized by thicker lines, are proportional to $N(I_1)$, indicating a higher density of CM channels. b) Schematic of intra-valley CM channel near $K$ ($K'$) points in monolayer $MoSe_2$. The highest valence band is depicted in gray, while the second lowest conduction band is shown in blue. An initial $2E_g$ -excited hot carrier within the second lowest conduction band is represented by a red solid circle, marked with the$|I_1\rangle$ state symbol. Near the $K$ ($K'$) points, the energy bands are nearly evenly spaced by $E_g$. The dashed and solid arrows in blue and red denote energy-momentum vectors that are parallel but in opposite directions, ensuring the conservation of energy and momentum. The empty circles represent the final carriers generated through this CM process, indicating the outcome of the carrier multiplication. c) $2E_g$ band nesting in monolayer $MoSe_2$. The $2E_g$ -shifted valence bands are depicted in blue. Regions of high intensity, highlighted in red, illustrate the substantial overlap between the $2E_g$ -shifted valence bands and the conduction bands. The highest valence bands and the second lowest conduction bands display similar energy-momentum, which we refer to as $2E_g$ band nesting.

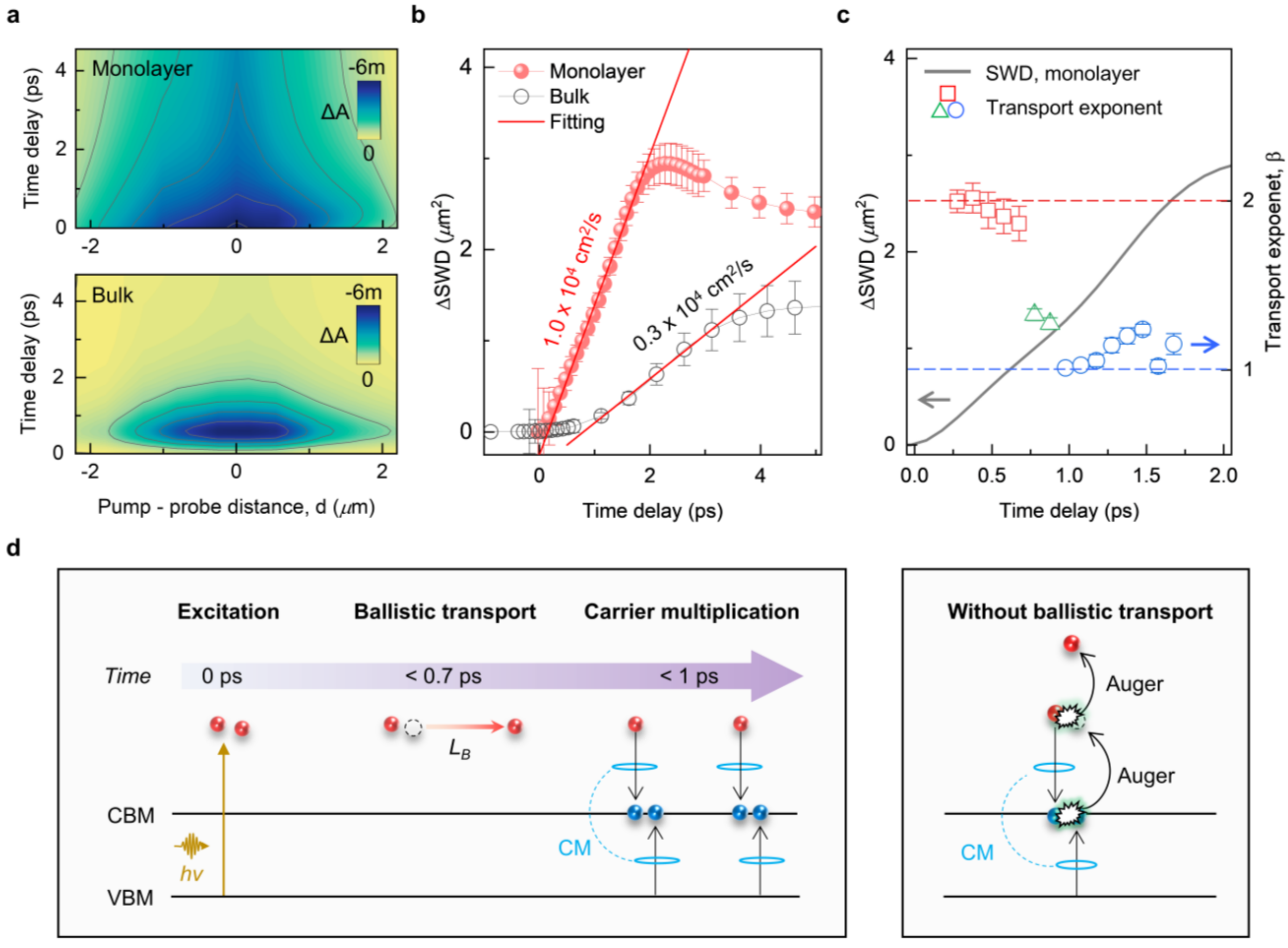

**Figure 3. Hot-carrier diffusion in monolayer and bulk $MoSe_2$ and impact of ballistic diffusion on carrier multiplication.**

a) Spatiotemporal evolution of hot-carrier distribution was investigated using transient absorption microscopy (TAM) in monolayer (**top**) and bulk (**bottom**). Each point is profiled at a specific time delay and pump-probe distance. The solid lines represent Gaussian fits (see **Note 7** in **Supporting information**). b) Initial hot-carrier diffusion is described by the differential square width distribution (ΔSWD) through Gaussian fitting to the spatial distribution as a function of time delay. The diffusion coefficient for monolayer and bulk, derived from linear fitting, are depicted**.** c) Temporal evolution of SWD (solid line) and the corresponding transport exponent, $\beta$ for monolayer. The transition from ballistic ($\beta$ = 2, red dot) to linear diffusion ($\beta$=1, blue dot) is marked. d) Schematic illustration of the role of ballistic transport during CM process. Upon photoexcitation, the carriers undergo quasi-ballistic transport spontaneously, leading to rapid spatial separation over a ballistic length ($L_B$). This spatial separation reduces the likelihood of Auger recombination, enhancing CM efficiency. In contrast, without ballistic transport, carriers remain in closer proximity, increasing the probability of Auger recombination and reducing CM efficiency.

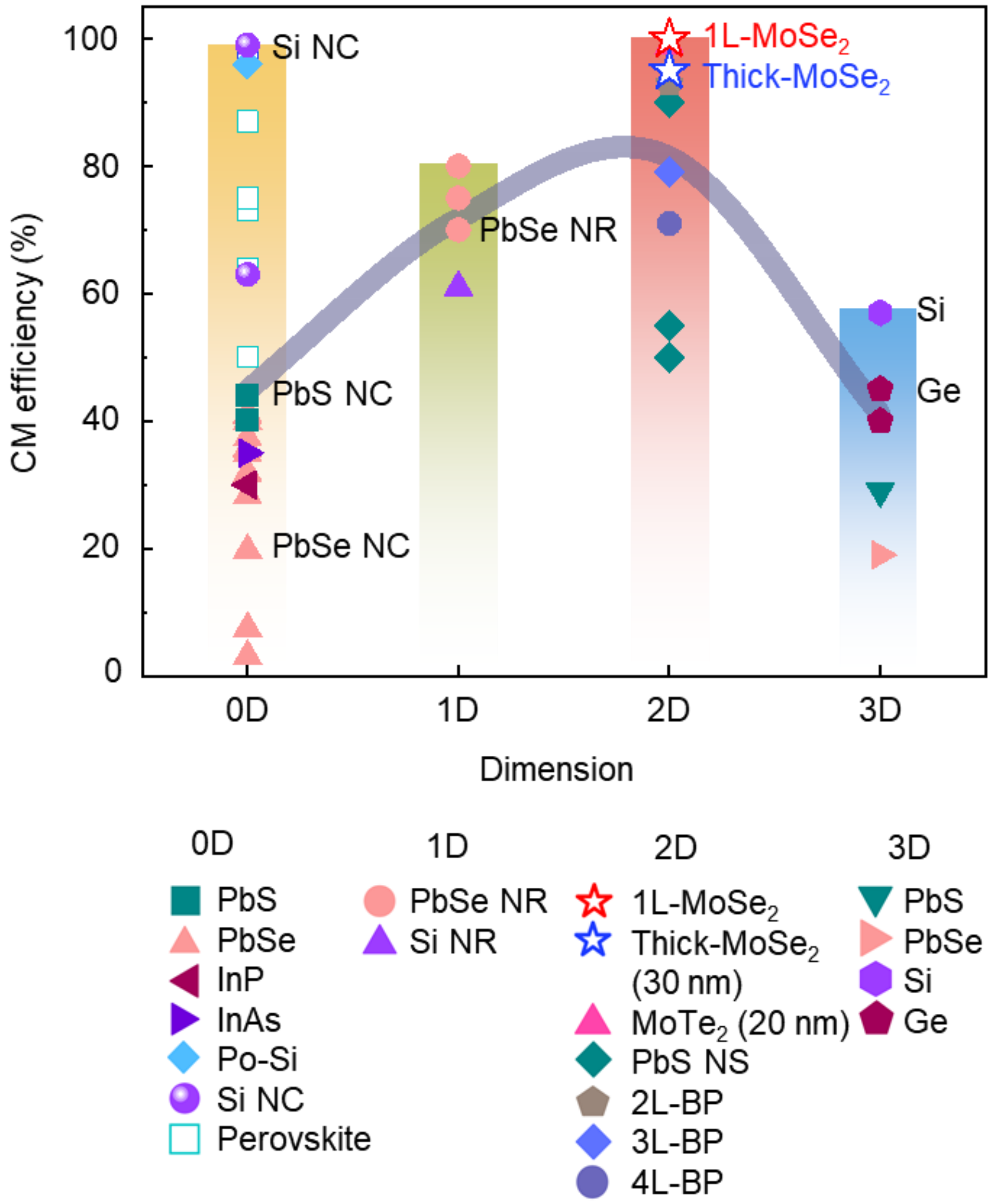

**Figure 4. Carrier multiplication conversion efficiency by dimensionality.** The CM efficiency is shown for a wide range of materials, categorized by dimensionality (refer to **Table S4** in **Supporting information** for detailed data and references). Each point represents the specific CM efficiency of a material, with vertical bars indicating the maximum values and horizontal line denoting the average values within each dimensional category.